\documentclass[showpacs,preprintnumbers,amsmath,amssymb,prb]{revtex4}

\usepackage{graphicx}
\usepackage{dcolumn}
\usepackage{bm}
\newcommand{\m}{\mathbf}
\newcommand{\h}{\hat}
\newcommand{\T}{\textbf}
\begin{document}

\title{A-phase origin in B20 helimagnets}

\author{S. V. Maleyev}
\affiliation{Petersburg Nuclear Physics Institute, Gatchina, St.\ Petersburg 188300, Russia}

\date{\today}

\begin{abstract}
 A-phase origin inn cubic helimagnets ($M n S i,F e G e$ etc) is explained. It is shown that its upper bound is a result of the spin-wave instability at $H_\perp>\sqrt{8/3}\Delta=H_{A2}$ where $H_\perp$ are the magnetic field perpendicular to the helix axis $\m k$ and $\Delta$  the spin-wave gap respectively. The last appears due to the spin-wave interaction if one takes into account that the Dzyaloshinskii-Moriya interaction acts between different spins. The infra-red divergences (IRD) in the $1/S$ series for the magnetic energy at $H_\perp\to H_{A2}$ are responsible for the lower A-phase boundary $H_{A1}$. It is shown that the A-phase exists at all $T<T_C$ but if $T\ll T_C$ it is very narrow and can not be observed. However in the critical region just below $T_C$ its width increasing strongly. Preliminary         estimations demonstrate semi-quantitative              agreement with  the existing experimental data.

 The existence of the spin-wave gap is a crucial point of our consideration. In Appendix A we present its derivation in more transparent form than in previous publication.
\end{abstract}

\pacs{61.12.B1,71.15.Rf,}

\maketitle

 \section{Introduction}
   
Unusual properties of noncentrosymmetric cubic B20 helimagnets ($M n Si,\; F e G e$ and related compounds) with
 Dzyaloshinskii-Moriya interaction (DMI) attracted a lot of attention during more than thirty years (see for example \cite{P,M,G1,G2}  and references therein).  Renascence in this field began with a discovery of a quantum phase transition to a disordered (partially ordered) state in $Mn Si$  at high pressure \cite{P1,K,P2,P3}. Then this transition was observed in $F e G e$ also \cite{PE}.
 
 Apparently complicated behavior of the helix wave-vector $\m k$ (the helix axis) in magnetic field $\m H$ is one of the most striking phenomenon observed in these compounds. Indeed   suppressing the weak cubic anisotropy the field  aligns  the helix axis along itself and gives rise  a conical magnetic structure which transforms to the ferromagnetic one at critical field $H_C$. This simple behavior holds almost in whole region of the  $(H,T)$ phase diagram. However just below the transition $T_C$ to paramagnetic state 
 the helix vector $\m k$  suddenly rotates perpendicular to the field. This so-called A-phase discovered by B. Lebech in $F e G e$ \cite{L} exists in rather narrow field range
  $H_{A1}<H<H_{A2}<H_C$ and above $H_{C2}$ the helix axis returns to the field direction  again
  \cite{L,P,M,G3}. 
 
Further small angle neutron scattering experiments revealed unexpected feature. If the neutron beam is  along the field the small-angle magnetic scattering appears in the A-Phase only  and
the magnetic Bragg reflections form the six-fold structure which dependent weakly on the field orientation relative to the crystal axes. This phenomenon was observed in $M n S i$ \cite{M}, $Fe G e$ \cite{G4} and several B20 compounds \cite{P}. It could be described  as superposition of three helices with wave vectors $\m k_1+\m k_2+\m k_3=0$  where $|\m k_i|=k$ which are perpendicular to the  field $\m H$.

 Meanwhile it was claimed that this structure is a  new skyrmion lattice state  \cite{M,P}. However the higher-order reflections which has to be in this case were not observed. Hence a nature of this six-fold structure inside the-A  has not been understood yet.

Moreover the very existence of the A-phase with perpendicular $\m k$ orientation has to be explained. Indeed  in zero field the multi-domain state is realized with $\m k$-axes are along $\langle 111\rangle$ or $\langle 001\rangle$ depending on a sign of very weak cubic anisotropy \cite{B}. Then with the field increasing the vectors $\m k$ rotate to the field and conical helix structure develops with the cone angle determined by $\sin\alpha=-H/H_C$ \cite{M1,G5}. This behavior is  the same as in antiferromagnets above   the spin-flop  transition where we have not any hint to the A-phase state. It should be noted also that the classical magnetic energy depends on $\m H_\parallel$ the field component along the helix vector $\m k$ only and $\m H_\perp$ dependence appears as a quantum phenomenon only \cite{M1}.

  Behavior of  the cubic helimagnets in the field is related with  more important multiferroic problem. Indeed in $R M n O_3$ materials multiferroic properties are connected directly to spin helices mediated by  DMI \cite{K1,GO,N}  and a lot of transitions in magnetic field were observed (see \cite{A} and references therein). However we have not now any  satisfied explanation of these transitions. Hence  understanding of the helices  behavior in magnetic field may be considered  as an  urgent problem.

In this paper we explain the A-phase origin  developing  theory of the helix behavior in the field which takes into account the  interaction between spin-waves.   
 This interaction gives rise the infra-red divergences (IRD) in perturbation $1/S$ series for the field-depending part of the magnetic  energy at $H^2_\perp-H^2_{A2}\to 0$, where $\m H_\perp$ is the field component perpendicular to the helix vector $\m k$ and $H_{A2}$ is the upper A-phase boundary respectively. We demonstrate also that $H_{A2}=\sqrt{8/3}\Delta$ where $\Delta$ is the spin-wave gap.  In the Hartree-Fock approximation it is given by \cite{M1}
\begin{equation}
	\Delta^2_{HF}=\frac{(A k^2)^2}{4S}\sum\frac{D_\m q}{D_0},
\end{equation}
where $S$ is the unit-cell spin. 

We demonstrate that the magnetic energy is a sum of two parts $E_M=E_\parallel+E_\perp$ differently  depending on $H_\parallel$ and $H_\perp$ as shown in Fig.1. This anisotropy is a consequence of an unusual form of the  spin-wave energy at $H_\parallel\ll H_C$\cite{M1,BC}
 \begin{equation}
 \begin{split}
	\epsilon_\m q&=
	\sqrt{(A k)^2(q^2_\parallel+3q^4_\perp/8k^2)+\Delta^2-3H^2_\perp/8},\;q\leq k;\\
	&A q\sqrt{q^2+k^2},\; q\gg k,
	\end{split}
\end{equation}
where $A$ is the spin-wave stiffness at $q\gg k$. This $\m q$ anisotropy is a result of the DMI which  mixes  spin-waves with $\m q$  and $\m{q\pm k}$ \cite{M1,B1}. These umklapps produce  satellite spin-wave structure  with $\m q\pm n \m k$ where $n=1,2,...$  observed  in \cite{J}.

\begin{figure}
\centering
\includegraphics{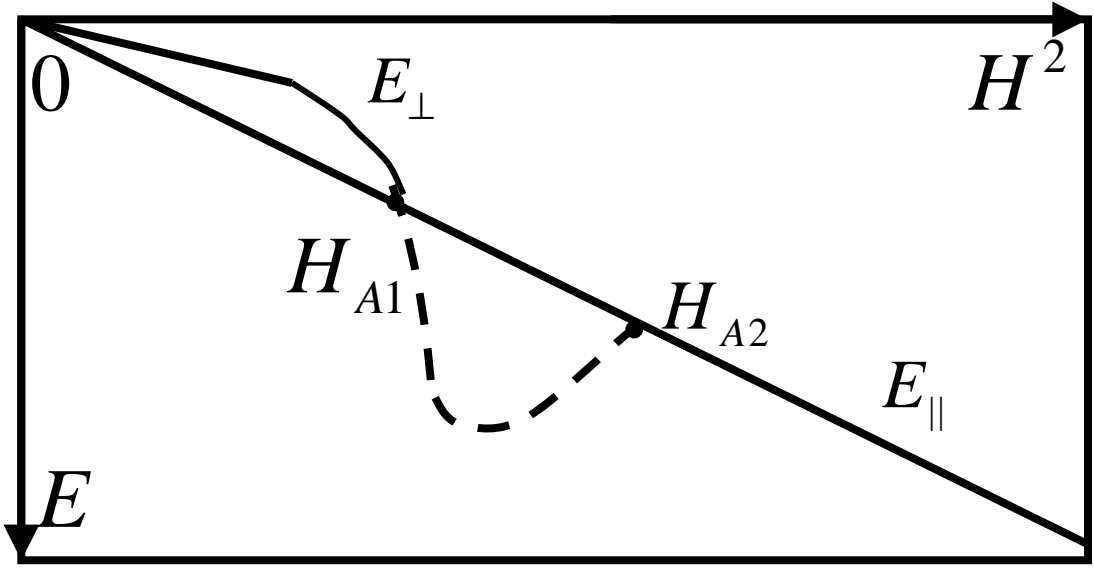}
\caption{Two parts of the magnetic energy   $E_\parallel$ and   $E_\perp$ with the the helix vector $\m k$  along and perpendicular to the field respectively. Between $H_{A1}$ and $H_{A2}$ in the A-phase region present approach is applicable in the multi-domain case only.}
\normalfont\tiny\rmfamily\scriptsize\mdseries\tiny
\label{fig.1}
\end{figure}

It should be noted that
 in \cite{B1} was claimed that in B20 helimagnets the spin-waves are the gapless Goldstones due to translation invariance along the helix vector $\m k$. Really this statement is correct in the linear theory only and is broken in higher approximations if one takes into account that the DMI acts between different 
spins \cite{M1,M4,M2,M3}. The essence of the problem is following. Conventional macroscopic expression for the DMI energy is given by $D\m{(M(r)\cdot[\nabla\times M(r)])}$ is ill-defined as both $\m M$ operators act in the same $\m r$ point. It is unimportant in the cases of the classical ground0state energy and the linear spin-wave theory but becomes crucial if one is interested in the spin-wave interaction. Indeed if both $\m M$ operators act in the same point they non-commute and we have not a gap. Otherwise they commute and the gap appears in the Hartree-Fock approximation \cite{M1,M4}(see also line below Eq.(15) and Appendix A). In this case the DMI feels the lattice structure  the above mentioned translation invariance is broken.
  The sum in Eq.(1) for $\Delta^2$ is saturated at $q\sim 1/a$ where the macroscopic approach is not applicable. Moreover in \cite{M2}  was shown that
the  magneto-elastic interaction  mixes $\m k=0$ magnons with the $2\m k$ phonons. As a result in the second order on this interaction a negative contribution to $\Delta^2$ appears 
and  we have
\begin{equation}
	\Delta^2=\Delta^2_{HF}+\Delta^2_{ME},
\end{equation}
 It was suggested \cite{M2,M3} that the quantum phase transition at pressure observed in \cite{P1,K,PE} is a result of a competition between these two terms and holds when $\Delta^2=0$.

From Eq.(2) follows also that uniform spin-waves are unstable if
\begin{equation}
	H_\perp>H_{A2}=\Delta\sqrt{8/3},
\end{equation}
and the perpendicular state is impossible. Hence $H_{A2}$ is the upper A-phase bound. Below in agreement with experiment we assume that $H^2_{A2}< H^2_C$ and demonstrate  that the lower bound $H_{A1}$ is a result of the infra-red divergences (IRD) which appear in the $1/S$ perturbation expansion for the magnetic energy if $H_\perp\to H_{A2}$. At low $T$ this boundary is very close to $H_{A2}$ and the A-phase can not be seen. However just below the transition $H_{A1}$ decreases strongly due to critical slowing down (decreasing of the spin-wave stiffness $A$ as $T\to T_C$)  and the A-phase becomes visible. Qualitative agreement with experimental data of Ref.\cite{M} was demonstrated. For detailed comparison with experiment one has to have more precise experimental data for the $T$ dependence of $H_{A1,2}$ in the critical region. 

\section{Spin-wave interaction}

  At the beginning  we have to summarize briefly principal theoretical results which will be explored below.  We use the Bak-Jensen model \cite{B} adding the Zeeman energy and omitting  weak cubic anisotropy which is unimportant. Corresponding Hamiltonian is given by
\begin{equation}
	H=\sum\{-J_\m q\m{(S_q\cdot S_{-q})}/2+i D_\m q\m{(q\cdot[S_q\times S_{-q}])}+\sqrt{N}\m{(H\cdot S_0)}\},
	\end{equation}
 where the first second and third terms are the ferromagnetic exchange interaction, DMI and the Zeeman energy respectively. In the case of the helical structure for $\m{S_q}$ we have\cite{M1}
\begin{equation}
	\begin{aligned}
	\m{S_q}&=S^c_\m q \hat c+S^A_\m q \m A+S_\m q^{A^*} \m A^*\\ 
	S^c_\m q&=S^\zeta_\m q  \sin\alpha+S^\xi _\m q  \cos\alpha,\:
	S^A_\m q=S^\zeta_\m{q-k}\cos\alpha-S^\xi_\m{q-k}\sin\alpha+i S^\eta_\m{q-k},\:
	S^{A^*} _\m q=S^\zeta_\m{q+k}\cos\alpha-S^\xi_\m{q+k}\sin\alpha-i S^\eta_\m{q+k},
	\end{aligned}
\end{equation}
	where $\m A=(\h a-i\h b)/2$, unit vectors $\h a, \h b$, $\h c$ form the right-handed orthogonal frame   and  the spin operators in the Dyson-Maleyev representation are given by
	\begin{equation}
	S^\zeta_\m q=N^{1/2}S\delta_{\m q,0}-(a^+a)_\m q;\:S^\eta_\m q=-i\sqrt{S/2}[a_\m q-a^+_{-\m q}-(a^+a^2)_\m q/2S];\:S^\xi=\sqrt{S/2}[a_\m q+a^+_{-\m q}-(a^+a^2)_\m q/2S],
\end{equation}
where ~$a_\m q \:\mbox{and}\:a^+_\m q$ ~are conventional  Bose  operators. 
 
In the classical approximation we have 	
	 $\m k=S D_0\h c/A$ where $A=S(J_0-J_\m k)/k^2$  is the spin-wave stiffness at $q\gg k$,  $\sin\alpha=-H_\parallel/H_C$ where $H_\parallel$ is the field component along $\m k$ and $H_C=A k^2$ \cite{B,M1}. Corresponding part of the magnetic energy have the form $E(H_\parallel)=-H^2_\parallel/2H_C$.
	  
	  We are interested below low-field region $H<H_C$ and for simplicity put $\alpha=0$. 	 Conventional spin-wave Hamiltonian is given by  \cite{M1}
\begin{equation}
	H_2=\sum[E_\m q a^+_\m q a_\m q+B_\m q(a_\m q a_\m{-q}+a^+_\m{-q}a^+_\m q)/2],
\end{equation}
 where at $H_\parallel\ll H_C$ we have \cite{M1}
 \begin{equation}
  E_\m q=S(M_{0,\m k}-M_\m{q,k})+B_\m q\simeq A(q^2+k^2/2) , B_\m q=S(M_\m{q,k}-J_\m q)/2\simeq A k^2/2,    
  M_\m{q, k}=(J_\m{q+k}+J_\m{q-k})/2+2D_\m q(\m k\cdot \hat c),
  \end{equation} 
 where approximate equalities hold at $q\ll 1/a$ and $a$ is the lattice spacing. The equilibrium condition is given by $D_0(\m k\cdot\hat c)=A k^2/S$.
  
 The second term in Eq.(5) mixes excitations with $\m q$ and $\m{q\pm k}$  and  gives rise the $\m q$ anisotropy in Eq.(2) at $q\leq k$ \cite{M1,B1}. 
   
For following we has to consider the spin-wave interaction in more details  in comparison with \cite{M1}. From Eqs.(5-7) at small $\alpha$ for the interaction energy we have 
$V_I=V_4+V_6$  and
\begin{eqnarray}
	V_4&=&-(1/2)\sum (M_\m{1-3,k}-M_\m{1,k})a^+_\m 1a^+_\m{-1+2+3}a_\m 2 a_\m 3+(1/4)\sum (J_\m 1-M_\m{1,k})(a_\m 1+a^+_\m{-1})(a^+a^2)_\m{-1},\\ 
	V_6&=&-(1/16S)\sum(J_\m 1-M_\m{1,k})(a^+a^2)_\m 1(a^+a^2)_\m{-1}\simeq(A k^2/16S^2)\sum (a^+a^2)_\m 1(a^+a^2)_\m{-1}, 
	\end{eqnarray}
	where $\m{1,2,3}$ label corresponding momenta. 
  The first term in Eq.(10) is a generalization of the Dyson interaction for ferromagnets \cite{D}. We do not consider here the $V_6$ interaction as it gives further terms in  $1/S$ expansion. 
  
  In the Hartree-Fock (HF) approximation Eq.(10) gives rise non-Hermitian spin-wave Hamiltonian
  \begin{equation}
   H_{SW}=\sum[\tilde E_\m q a^+_\m q a_\m q+(\tilde B^+_\m q a_\m q a_\m{-q}+\tilde B_\m q a^+_\m{-q}a^+_\m q)/2],
   \end{equation}
where $\tilde E_\m q=E_\m q+\Sigma,\;\tilde B^{(+)}_\m q=B_\m q+\Pi^{(+)}$ and corresponding diagram is shown in Fig.2a. Corresponding expressions at $\m q=0$ are given by
 \begin{eqnarray}
	 \Sigma=(1/2)\sum[(J_\m 1-M_\m{1,k})(n_\m 1+f_\m 1)+(J_0-M_{0\m k})(n_\m 1+f_\m 1/2)];\\\nonumber 
	\Pi=\sum[(M_\m{0,k}-M_{\m 1,\m k})+\frac{1}{2}(J_0-M_{0,\m k})]f_\m 1;\\\nonumber
	\Pi^+=(1/2)\sum[(J_\m 1-M_\m{1,k})(n_\m 1+f_\m 1)+2(J_0-M_{0,k})n_\m 1].
\end{eqnarray}
	  where
	   \begin{equation}
	    n_\m 1=<a^+_\m 1 a_\m 1>=[(2N_\m 1+1)E_\m 1-\epsilon_\m 1]/2\epsilon_\m 1,\; f_\m 1=-B_\m 1(2N_\m 1+1)/2\epsilon _\m 1
	    \end{equation}
	     where $N_\m 1$ is the Plank function.   
	    From these expressions using definitions (9) for the spin-wave  gap we obtain  (for discussion see Appendix A)
\begin{equation}
\Delta^2_{HF}=2E_0\Sigma-B_0(\Pi+\Pi^+)=\frac{(A k^2)^2}{4S}\sum\frac{D_\m 1}{D_0},
\end{equation}
where $E_0=B_0=A k^2/2$.	     We have $\Delta_{HF}^2\neq 0$ due to $-1/2$ term in expression for $n_\m 1$. Forgetting that the DMI acts always between different spins we must replace $n_\m 1\to n_\m 1+1/2$ and $\Delta^2_{HF}\equiv 0$ \cite{M4}. It should be noted also that the sum in Eq.(13) is saturated at $q\sim 1/a$ and may be considered as $T$-independent at $T\leq T_C$.
 
 One has to note also that Fig.2a diagram is of order of $S^0$ and $\Delta^2\sim S$ whereas the main part of $\epsilon^2_\m q\sim S^2$.

	     \section{Perpendicular field ($\m{H\perp k}$)}  
     
From Eqs.(5-7) for interaction with the perpendicular field at $\alpha\ll 1$ we have \cite{M1}
\begin{equation}
	V_\perp=\m{(H \cdot A)}[\sqrt{S/2}(a_\m{-k}-a^+_\m k) 
	-\sum a^+_\m{q-k}a_\m q]+h.c.=V_L+V_{UM} ,
	\end{equation}
	where the  terms  linear in the $a^{(+)}_\m{\pm k}$ operators  lead to the spin-wave Bose condensation at $\m{q=\pm k}$. Other terms mix excitations with $\m q$ and $\m{q\pm k}$.
	
	 in Appendix B we obtain a system of equations   
	for $a^{(+)}_0$ and $a^{(+)}_\m{\pm k}$ considering them as classical variables.  From their solution we have\cite{NOTE}
\begin{equation}
	 a_\m k=\sqrt{S/2}\m{(H\cdot A)}/A k^2; a_\m{-k}=-\sqrt{S/2}\m{(H\cdot A^*)}/A k^2.
	\end{equation}
	   and the magnetic energy is given by \cite{NOTE} 
\begin{equation}
E_{M0}=-\frac{S H^2_\parallel}{2H_C}-\frac{S H^2_\perp}{4H_C}.
\end{equation}
where the factor $1/4$ is very transparent: $1/4=1/2<\cos^2\varphi>$ where $\varphi$ is the angle between the helical spin and  $\m H_\perp$.
According to this equation the helix axis $\m k$ has to be along the field as was observed in all $H,T$ region of the phase diagram except small A-phase pocket just below $T_C$ \cite{L,P,M,G3}.

 Taking into account the spin-wave BC we get from $V_4$ additional contribution to the spin-wave Hamiltonian (12)
  \begin{equation} 
  H_{B C}=-(H^2_\perp/16A k^2)\sum(a^+_\m q a_\m q+a^+_\m q a^+_\m{-q}+2a_\m q a_\m{-q}).
 \end{equation}
 As a result we obtained $-3H^2_\perp/8$ term in Eq.(1) instead of $-H^2_\perp/2 $  in \cite{M1}
 This part of the spin-wave Hamiltonian has been taken into account in Appendix B for the Green function evaluating    and produce non-symmetric $H^2_\perp$ terms in $F_\m q$ and $F^+_\m q$ Green functions (see below).
  
	 These umklapps gives rise to two infinite systems of conjugated   linear    equations for the Green functions. At $H_\perp\ll H_C$ they can be truncated  and we have two systems of the linear equations considered in Appendix. Their solution  may be divided on two parts: direct and umklapp Green functions. For the first one we have
	   \begin{equation}
	G_\m q(\omega)=\bar G_\m q(-\omega)=-\frac{i\omega+E_\m q+\Sigma-7H^2_\perp/16A k^2 }{\omega^2_n+\epsilon^2_\m q};\quad F_\m q=\frac{B_\m q+\Pi}{\omega^2_n+\epsilon^2_\m q};\quad F^+_\m q =\frac{B_\m q+\Pi^+-H^2_\perp/8A k^2}{\omega^2_\m q+\epsilon^2_\m q}.
	\end{equation}
	 The umklapp functions are given by
	 \begin{equation}
	 G_+=\bar G_-=-F^+_+=-F_-=-\frac{\m{(H\cdot A^)}}{2(\omega^2_n+\epsilon^2_\m q)};\quad  
	 G_-=\bar G_+=-F^+_-=-F_+
	 =-\frac{\m{(H\cdot A^*)}}{2(\omega^2_n+\epsilon^2_\m q)}, 
\end{equation}
where $\epsilon^2_\m q$ is given by Eq.(2)and in numerators we neglected some terms proportional to $q^2_\perp$ and $q^4_\perp$ which are  unimportant to us (see \cite{M1}).    
Expressions (14) for $n_\m q$ and $f_\m q$ are direct consequences of these equations.

  \begin{figure}
\centering
\includegraphics{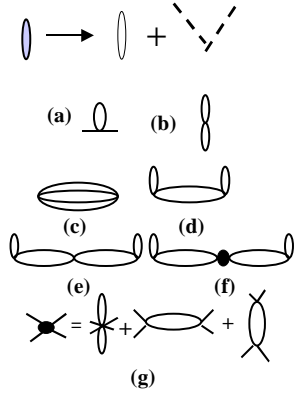}
\caption{ Diagrammatic expansion for the magnetic energy. Lines are  Green functions $G=\leftarrow$, $F=\rightarrow\leftarrow$ and $F^+=\leftarrow\rightarrow$. At top is the sum of the HF loop and the BC part Eq.(19) used in the diagrams. The IRD first time appears in Fig.2d diagram. 
 \label{fig.2}}
\end{figure}
  
\section{Magnetic energy}
 
     We demonstrate below that the A-phase lower bound is a result of the infrared divergence (IRD) at $H_{A2}-H_\perp\to 0$ which appear in the $1/S$ perturbation series for the interaction energy.   Using Eqs.(12, 18,19) for the total magnetic energy we have
 \begin{equation}
	E=E_{M0}+<H_{SW}+H_{BC}>+<V-V_{HF}>,
\end{equation} 
where the first term us given by Eq.(18). 
For the second term we have very transparent expression 
\begin{equation}	
	E_{SW}=\sum[\epsilon_\m q N_\m q+(\epsilon_\m q-E_\m q)/2], 
	\end{equation} 
	where off-diagonal  terms in Eq.(16) are responsible for correct $-3H^2_\perp/8$  field dependence of $\epsilon^2_\m q$. This expression includes the energy of thermally excited	magnons and zero-point motion with energy given by Eq.(2). The last term of Eq.(22) does not contain the HF part Fig.2b as   it was used in Eq.(12) ant the $1/S$ expansion begins with Fig.2b diagram. 

  In the principal $1/S$ order left and right vertexes in Fig.2 IRD diagrams may be expressed 
   using Eqs.(13) 
\begin{equation}
\Gamma=[\Sigma-(\Pi+\Pi^+)/2]a^+_\m q a_\m q+(1/2)(\Pi a^+_\m q+\Pi^+ a_\m{-q})(a_\m q+a^+_\m{-q}),
\end{equation}
 where  according to Eqs.(19,B5)we have to add to $\Sigma,\Pi^{(+)}$ the BC parts (fir st line in Fig.2).    . As a result we get
\begin{equation}
	\Gamma=(\Delta^2/A k^2)[ R_1a^+_\m q a_\m q-(R_2 a^+_\m q+R_3 a_{-q})(a_\m q+a^+_\m q)/2)]
\end{equation}
where $R_1=1+H^2_\perp/3H^2_{A2}\simeq 4/3,\;R_2=1+H^2_\perp/6H^2_{A2}\simeq 7/6,\mbox{and}\; R_3=H^2_\perp/3H^2_{A2}\simeq 1/3$. Approximate equalities  hold et $H_\perp\to H_{A2}$ and will be used  below.

Using Eq.(25) and taking into account  direct (20) and umklapp (21) Green functions for the first IRD diagram Fig.2d we obtain   
  \begin{equation}
	 E_d=-\frac{H^4_{A2} T}{8}\sum_{\m q,\omega_n}\frac{1+H^2_\perp/32A k^2 } {(\omega^2_n+\epsilon^2_\m q)^2},
	\end{equation}
	where we neglected in the numerator terms bilinear in  $(H^2_\perp, \Delta^2)$ and in following will omit the $H^2_\perp$ term.
	
	We are interested in $T\gg A k^2$
region and the IRD displaces the $\omega_n=0$ term only   and we obtain
	  \begin{equation}
	 E_d=-\frac{T(a k)^3\sqrt{3/8}H^2_{A2}}{18\pi(A k^2)^2}\left(\frac{H^2_\perp}{H^2_{A2}-H^2_\perp}+1\right)\simeq E_d\frac{H^2_\perp}{H^2_{A2}},
\end{equation}
   where $a$ is the lattice constant. and  instead of Eq.(18) we have
 \begin{equation}
	E_M= -\frac{S H^2_\parallel}{2H_C}-\frac{S H^2_\perp}{4H_C}\left(1-\frac{4H_C E_d}{S H^2_{A2}} \right)
\end{equation}
  Neglecting higher order IRD terms (see below) we obtain that the A-phase is energetically profitable if the expression in the  brackets becomes larger than two. Hence at the S-phase lower boundary we have
\begin{equation}
	E_d=-4S H^2_{A2}/H_C
\end{equation}
 and using Ea.(27)we get
 \begin{equation}
H_{A1}=H_{A2}[1-2\sqrt{3/8}(k a)^3T H_C/9\pi S (A k^2)^2 ]^{1/2}
\end{equation} 
  Using well known low-$T$ parameters for $M n Si$: $T_C\simeq 29.5K$, $H_C\simeq A k^2\simeq 0.6 T\simeq 0.8 K,\, a\simeq 0.46 nm,\, k\simeq 0.38 nm^{-1}$ and $S=1.6$(see for example \cite{M1} and references therein) we obtain $H_{A1}/H_{A2}=(1-0.0053 T/T_C)^{1/2}$. Hence at $T\ll T_C$  the A-phase exists but it is so narrow that can be hardly  observable. Moreover at $T=T_C$ the ratio $r=H_{A1}/H_{A2}=0.9975$.
 
 We demonstrate now that 
   there is strong decreasing of the ratio $r=H_{A1}/H_{A2}$ just below $T_C$. It is connected mainly with the spin-waves critical slowing down.
  Unfortunately there are not any theoretical predictions for  it and our discussion is restricted by analysis of
  Eq.(30) where all parameters may be measured independently.
  
 First of all the low-$T$ condition $H_C\simeq A k^2$ is violated. Indeed $H_C(T_C)\sim H_C(0)/2$ \cite{M}, $k$ is almost $T$ independent  \cite{G5} whereas
 the spin-wave stiffness $A$ is renormalized strongly. Indeed it is equal to $0.52meV nm^2$ and $0.24meV nm^2$  at $T=5K$ and $26K$  respectively \cite{I,S,CM}  while  $H_C$ remains almost unchanged \cite{M}.  One may await  that near $T_C$ $A=A_0 \tau^z$ where $\tau=(T_C-T)/T_C$.
 
  The unit cell spin $S=S_0 \tau^{0.22}$ \cite{GE,G6}. According to \cite{G7} this scaling behavior begins at $T\approx 25K\;(\tau\approx 0.14)$ and $S_0\approx 1.2$. 
  The upper A-phase bound is  almost $\tau$ independent \cite{M,G3} and   from Eqs.(1) and (4) we have $A\sim S^{1/2}$ and $A=A_0\tau^{0.11}$. Putting $A_0=0.24meV nm^2$ we obtain
\begin{equation}
	H_{A1}=H_{A2}(1-0.033 h/\tau^{0.44})^{1/2},
\end{equation}
 where $h=H_C(T)/H_C(0)$. 
 
The most detailed available data for the A-phase in $M n S i$ are shown in Fig.1 of Ref.\cite{M} and we compare our results with them. i. $T=25K,\;\tau=0.14,\;h\simeq 1\; r=H_{A1}/H_{A2}=0.96$ and A-phase was not seen. ii. $T\simeq 28.5K,\;\tau=0.03\;h\simeq 0.8$;  $r=0.94$ whereas the observed ratio  is  $ 0.6-0.7$. Hence the theory explains the A-phase phenomenon at least qualitatively. It has to mention also that all parameters including values of $\tau$ used above are known with very low  accuracy. For example at $\tau=0.01$  we have
$r=0.89$ and replacing in Eq.(31) $0.033\to 0.1$ we get $r=0.63$ 
 We mention also that according to \cite{M,G5}
\begin{equation}
	H_{A2}\simeq 0.22T;\;\mbox{and}\;\Delta\simeq 0.13T=15\mu e V.
	\end{equation}
 
 \section{$1/S$ corrections}
 
The next IRD term is represented by Fig.2e diagram and in general form is given by
\begin{equation}
	E_e=<\Gamma V_4\Gamma>.
\end{equation}
  Interaction $V_4$ given by Eq.(10) consists of two parts.   The first one gives  the central vertex in Fig.2e diagram which disappears at zero momenta and 
 leads to rather weak IRD 
 singularity  which may be neglected. The second remains constant and gives main correction to the above results.

 Using Eqs.(2),(20) and (A3) for the most singular  $e2$ correction we obtain 
\begin{equation}
	E_e=-\frac{3A k^2}{S}\left(\frac{\Delta^2 R_1}{A k^1}\right)^2\sum\frac{B_\m 1(F^+_\m 1-F_\m 1)B^2_\m 2}{Z_\m 1 Z^2_\m 2}\to \frac{15E_d A k^2 H^2_\perp}{32S}\sum\frac{T}{\epsilon^4_\m 1},
\end{equation}
   where $Z_l=\omega^2_{n_l}+\epsilon^2_\m l$ and $l=1,2$. This expression is proportional to $(H^2_{A2}-H^2_\perp)^{-2}$ Using Eq.(26) we obtain $\sum(T/\epsilon^4)=-8E_d/H^4_{A2}$ and 
\begin{equation}
	E_e=-\frac{15 E^2_d A k^2 H^2_\perp}{4 S H^4_{A2}}
\end{equation}
Considering this expression as a small correction to $E_d$  in Eq.(29) we must replace
\begin{equation}
	E_d\to E_d(1+15Ak^2/16H_C).
\end{equation}
At low $T$ this correction of order of unity and one has to examine all $2/S$ series. However near $T_C$ where $A\to A_0\tau^{0.11}$ it may be neglected. In any case deep into the A-phase and near its upper boundary the
 full series examining is unavoidable if one assume conventional helical structure (cf\cite{M}). 
  
   \section{Conclusions}
   
    A-phase origin in cubic helimagnets ($M n S i,F e G e$ etc) is examined. It is shown that its upper bound is a result of the spin-wave instability at $H_\perp>\sqrt{8/3}\Delta=H_{A2}$.
    
    The infra-red divergences (IRD) in the $1/S$ series for the magnetic energy at $H_\perp\to H_{A2}$ are responsible for the lower A-phase boundary $H_{A1}$. It is shown that these IRD are anomalously strong due to softens of the spin-wave spectrum at $q<k$. As a result the low-momenta fluctuations behave as in 2D systems \cite{KI}.

     We demonstrate that the A-phase exists at all $T<T_C$ but if $T\ll T_C$ it is very narrow and can not be observed. However in the critical region just below $T_C$ its width increasing strongly. Preliminary         estimations show semi-quantitative              agreement with  the existing experimental data. However the problem demands further theoretical and experimental studies. We can formulate following unresolved problems.
     i. Examination of the full $1/S$ expansion at low $T$ as the second correction to the $H_{A1}$ boundary is not small. In this respect we wish to point out to the hint to the A-phase observed in $M n S i$  at $T=10K$ (Fig.3b in Ref. \cite{G5}). ii. Better understanding the temperature dependence of the spin-wave energy at $q<k$. iii. More precise measurements of the A-phase boundaries then it was done in previous studies. In this respect we wish to note that there are the first order transitions at both a-phase boundaries which has to be accompanied by the specific heat jumps. The first time the specific heat anomalies in this region were observed in \cite{ST} but more precise measurements would be important. 
     
     In this discussion we avoided the nature of the A-phase itself. There are two possibilities: exotic "skyrmion" state proposed in \cite{M} or the tree domain structure. However at present we have not any real theory of the skyrmion state as well as an explanation of the   three domain state.
     
\begin{acknowledgments}
  
   This work was partly supported by RFBR grants 09-02-00229,
   10-02-01205-à, Goskontrakt No 02.740.11.0874.
and Programs "Quantum Macrophysics", "Strongly correlated electrons in semiconductors, metals, superconductors and magnetic materials" and "Neutron Research of Solids".
 
\end{acknowledgments}

    \appendix
 \section{} 
  
 As very existence of the gap is crucial for us
   we repeat here some details of corresponding calculations which were presented in \cite{M1} in a rather cumbersome form. In Eq.(15)  terms proportional to $J_0-M_\m{q,k}$ cancel. For remained parts of $\Sigma,\;\Pi^{(+)}$ using Eqs.(9,13,14) we obtain
\begin{eqnarray}
	\Sigma=\Pi^+;\;\Pi=\Sigma-\sum(M_\m{1,k}-J_\m 1)/4&\\
	 \sum(M_\m{1,k}-J_\m 1)/4=(1/2)\sum D_\m q (\m k\cdot \hat c)
	 \end{eqnarray}
as $\sum J_\m{1,k}=\sum J_\m q=0$  due to condition that the exchange interaction acts between different spins. Using equality $D_0(\m k\cdot \hat c)=A k^2/S$  we obtain Eq.(1). 

It may be shown that expression (A2) is much larger than other terms in Eqs.(13) as they have additional small factors which vanish if $\m k\to 0$. Hence we may put    
\begin{equation}
	\Pi=-2\Delta^2/A k^2,\;\Sigma=\Pi^+=0. 
\end{equation}
   
From Eq.(A2) follows that $\Delta^2\neq 0$ if $\sum D_\m q\neq 0$. We demonstrate now that this condition is fulfilled. 
In \cite{M1} for the DMI was suggested expression 
\begin{equation}
	V_{DM}=(1/)\sum D_{\m R,\m R'} (\nabla-\nabla')[\m S_\m R \times \m S_{\m R'}]=i\sum D_\m q (\m q\cdot\m{[S_q\times S_\m{-q}]}),
\end{equation}
where $D_\m{R,R'}=D_\m{R',R}$, the condition $\m{R\neq R'}$ holds
and $\sum\m q D_\m q$=0.

 We present now  simple example where the $\sum D_\m q\neq 0$ was evaluated.

In \cite{Y}  a simple model for the DMI in cubic lattice was proposed where the DMI acts between neighboring spins and the DMI vector $d_{j,\m q}=-i K \sin q_j,\;j=x,y,z.$. For the form-factors we have $(\sin q_j)/q_j$ and  $\sum \sin q_j/ q_j=\pi/2$. 

It should be  pointed again  the gap appears if   the DMI acts between different spin  and the translation invariance along $\m k$ is broken. Otherwise this invariance is restored and the gap disappears due to non-commutativity of two spin operators in the single lattice point \cite{M4}.

   \section{}
     At $H_\perp\ll H_C$ truncated equations for the Bose condensed magnons are given by (cf. \cite{M1})
   \begin{eqnarray}
  E_0 a^+_0+ B^+a_0-h a^+_\m k-f a^+_\m{-k}  &=&0,\\\nonumber
  B a^+_0+  E_0 a_0-h a_\m{-k}-f a_\m k=&0,\\\nonumber
-f a^+_0+E_\m 1 a^+_\m k+B a_\m{-k}=&\sqrt{S/2}f,\\\nonumber
-f a_0+B a^+_\m k+E_\m 1 a_\m{-k}=&-\sqrt{S/2}f,\\\nonumber
-h a^+_0+ E_\m 1 a^+_\m{-k}+B a_\m k=&-\sqrt{S/2}h,\\\nonumber
-h a_0+B a^+_\m{-k}+ E_\m 1a_\m k=&\sqrt{S/2}h,
\end{eqnarray}
and we obtain Eqs.(17).
   
 We have  two sets of the Green functions determined as follow
\begin{eqnarray}
	G_\m q&=&-<Ta_\m q,a^+_\m q>;F^+_\m q=-<Ta^+_\m{-q},a^+_\m q>;G_\pm=-<Ta_\m{q\pm k}, a^+_\m q>;F^+_\pm=-<Ta^+_{\m{-q\mp k}},a^+_\m q>,\\
	\bar G_\m q&=&-<Ta^+_\m q,a_\m q>;F_\m q=-<Ta_\m{-q},a_\m q>; \bar G_\pm=-<Ta^+_\m{q\pm k},a_\m q>;F_\pm=-<Ta^+_{\m{-q\mp k}},a_\m q>,
	\end{eqnarray}
	 They are solutions of two conjugated  systems of linear equations. W e consider here the second set only. Corresponding truncated system is given by
\begin{eqnarray}
(i\omega+\tilde E)\bar G+\tilde B^+ F-h \bar G_+-f \bar  G_-&=&-1,\\\nonumber
-\tilde B\bar G+(i\omega-\tilde E)F+h F_++f F_-=&0,\\\nonumber
-f\bar G+(i\omega+E_\m 1)\bar G_++B F_ +=&0,\\\nonumber
f F-B\bar G_++(i\omega-E_\m 1)F _+=&0,\\\nonumber
-h\bar G+(i\omega+E_\m 1)\bar G_-+B F_-=0,\\\nonumber
h F_--B \bar G_-+(i\omega-E_\m 1)=&0,
\end{eqnarray}
 where $h=f^*=(\m{A\cdot H})$, $E_\m 1\simeq E_\m{q\pm k}$  if $q\ll k$ and according to Eqs.(12,13) and the BC contribution (19) we have
\begin{equation}
\tilde E=E_\m q+\Sigma-H^2_\perp/16A k^2;\, \tilde B=B+\Pi-H^2_\perp/8A k^2;\, \tilde B^+=B+\Pi^+-H^2_\perp/4A k^2,
\end{equation}
  Solution of these equations as  well as the conjugated one are given by Eqs.(20) and (21).


\begin{thebibliography}{55}
 
\bibitem{P} C.Pfleiderer, C.Adams, A.Bauer et al., J.Phys.: Cond. Matter \T{22}, 164207 (2010),
\bibitem{M} S.M\"uhlbauer, B.Binz, F.Jonietz et al., Science \T{323}, 915 (2009),
\bibitem{G1} S.V.Grigoriev, D.Chernenkov, V.A.Dyadkin et al., Phys.Rev.Lett. \T{102}, 037204 (2009).
\bibitem{G2} S.V.Grigoriev, D.Chernysov, V.A.Dyadkin et al., Phys.Rev.B, \T{81}, 012408 (2010), 

\bibitem{P1} C.Pfleiderer, G.J.MacMillan, S.R.Julian, G.G.Lonzarich, Phys.Rev.B \textbf{55}, 8330 (1997).
\bibitem{K} K.Koyama, T.Goto, T.Kanomata, R.Note, Phys.Rev.B \textbf{62}, 986 (2000).

\bibitem{P2} C.Pfleiderer, S.R.Julian, G.G.Lonzarich, Nature(London) \textbf{414}, 427 (2001).
\bibitem{P3} C.Pfleiderer, D.Reznik, L.Pintschovius, H.v.L\"ohneysen, M.Garst, A.Rosh, Nature (London) \textbf{427}, 227 (2004).
\bibitem{PE} P.Pedrazzini, H.Wilhelm, D.Jaccard et al., Phys.Rev.Lett. \T{98}, 047104 (2007).
  \bibitem{L} B.Lebech, Recent Advances in Magnetism of Transition Metal Compounds (Word Scientific, Singapore) p.167 (1993).
  \bibitem{G3} S.V.Grigoriev, V.A.Dyadkin, E.V.Moskvin, D.Lamago, Th.Wolf, H.Eckerlebe and S.V.Maleyev, Phys.Rev.B \T{79}, 144417 (2009).
  \bibitem{G4} S.V.Grigoriev, S.V.Maleyev E.V.Moskvin, V.A.Dyadkin, P.Fouquet and H.Eckerlebe, Phys.Rev.B \T{81}, 144413 (2010). 
  \bibitem{B} P.Bak, M.Jensen, J.Phys.C  \textbf{13}, L881 (1980).
  \bibitem{M1} S.V.Maleyev, Phys.Rev.B \T{73}, 174402 (2006).
 \bibitem{G5} S.V.Grigoriev, S.V.Maleyev, A.I.Okorokov, Yu.O.Chetverikov, P.B\"oni, R.Georgii, D.Lamago, H.Eckerlebe, K.Pranzas, Phys.Rev.B  \textbf{74}, 214414 (2006).
  
 \bibitem{K1} T.Kimura, T.Goto, H.Shintani et al., Nature \T{426}, 55 (2003).
 \bibitem{GO} T.Goto, T.Kimura, G.Lawes et al., Phys.Rev.Lett. \T{92}, 257201 (2004).
 \bibitem{N} N.Nagaosa,J. Phys.: Condens. Matter \T{20}, 434207 (2008).
 \bibitem{A} N.Aliouane, K.Schmaltl, D.Senff et al, Phys.Rev.Lett \T{102}, 207205 (2009). 
 \bibitem{BC} In  Eq.(2) we have $-3H^2_\perp/8$ instead of $-H^2_\perp/2$ in \cite{M1} where the spin-wave Bose condensation in the spin-wave Hamiltonian was not be considered. See Eq.(19) below.
  
 \bibitem{B1}  D.Belitz, T.R.Kirpatrick and A.Rosch, Phys.Rev.B   \textbf{73}, 054431 (2006).
 
 \bibitem{J} J.Janoschek, f.Bernlochner, S.Dumsiger, C.Pfleiderer, P.B\"oni, D.Rossli , D.Link and A.Rosch, Phys.Rev.B  \textbf{81},214436 (2010).  
 \bibitem{M4} S.V.Maleyev, arXiv: 0711.3547,

\bibitem{M2} S.V.Maleyev, J.Phys.Condens.Matter \T{21}, 146001 (2009).
\bibitem{M3} S.V.Maleyev, JMMM \T{321}, 909 (2009).
 \bibitem{NOTE} In Refs. \cite{M1,M2,M3} expressions for $a_\m{\pm k}$ and the perpendicular part of the magnetic energy have addition factors  which really equal to unity.

\bibitem{D} F.J.Dyson, Phys.Rev. \T{102}, 1217; 1230 (1956).
  \bibitem{GE} R.Georgii, P.B\"oni, D.Lamago et al., Physica B \T{350}, 45 (2004).
\bibitem{G6} S.V.Grigoriev, S.V.Maleyev, A.I.Okorokov et al., Phys.Rev.B \T{72}, 134420 (2005).
\bibitem{G7} S.V.Grigoriev, private communication.

\bibitem{I} Y.Ishikawa, G.Shirane, J.A.Tanaka and M.Kogio, Phys.Rev.B \T{16}, 4956 (1977).
\bibitem{S} F.Semadeni, P.B\"oni, Y.Endoh, B.Roessly and G.Shirane, Physica B \T{267-268}, 248 (1999).
\bibitem{CM} Unfortunately the spin-wave stiffness was measured in "ferromagnetic" range where momentum transfer $q>k$ and we have not any direct information about the $A$ renormalization at $q\ll k$.
\bibitem{KI} T.R.Kirkpatrick and D.Belitz, Phys.Rev.Lett. \T{97}, 267205 (2006).
\bibitem{ST} S.M.Stishov, A.I.Petrova, S.Khasanov, G.Kh.Panova, A.A.Shikov. J.C.Lashley, D.Wu and I.A.Lagrasso, J.Phys.:Condensed Matter \T{20}, 235222 (2008). 
 \bibitem{Y} X.Z.Yu, Y.Onosem, N.Kanasawa, J.H.Park, J.H.Han, Y.Matsui, N.Nagaosa and Y.Tokura, Nature \T{465/17}, 90 [Supplementary Information] (2010) 


\end{thebibliography}
\end{document}